\documentclass{edm_article}
\usepackage{booktabs}
\begin{document}

\title{Examining the Role of LLM-Driven Interactions on Attention and Cognitive Engagement in Virtual Classrooms}


\numberofauthors{1}
\author{
\alignauthor Süleyman Özdel\textsuperscript{\normalsize 1}, 
Can Sarpkaya\textsuperscript{\normalsize 1}, 
Efe Bozkir\textsuperscript{\normalsize 1}, 
Hong Gao\textsuperscript{\normalsize 2}, 
Enkelejda Kasneci\textsuperscript{\normalsize 1} \\
\affaddr{\textsuperscript{1}Human-Centered Technologies for Learning, Technical University of Munich, Germany}\\
\affaddr{\textsuperscript{2}School of Future Science and Engineering, Soochow University, China}\\
\email{\{ozdelsuleyman, can.sarpkaya, efe.bozkir, enkelejda.kasneci\}@tum.de, gaohong@suda.edu.cn}
}

\maketitle

\begin{abstract}
Transforming educational technologies through the integration of large language models (LLMs) and virtual reality (VR) offers the potential for immersive and interactive learning experiences. However, the effects of LLMs on user engagement and attention in educational environments remain open questions. In this study, we utilized a fully LLM-driven virtual learning environment, where peers and teachers were LLM-driven, to examine how students behaved in such settings. Specifically, we investigate how peer question-asking behaviors influenced student engagement, attention, cognitive load, and learning outcomes and found that, in conditions where LLM-driven peer learners asked questions, students exhibited more targeted visual scanpaths, with their attention directed toward the learning content, particularly in complex subjects. Our results suggest that peer questions did not introduce extraneous cognitive load directly, as the cognitive load is strongly correlated with increased attention to the learning material. Considering these findings, we provide design recommendations for optimizing VR learning spaces.
\end{abstract}

\section{Introduction}
\label{sec:intro}

\begin{figure*}[ht]
    \centering
    \begin{minipage}{0.45\textwidth}
        \centering
        \includegraphics[width=\textwidth]{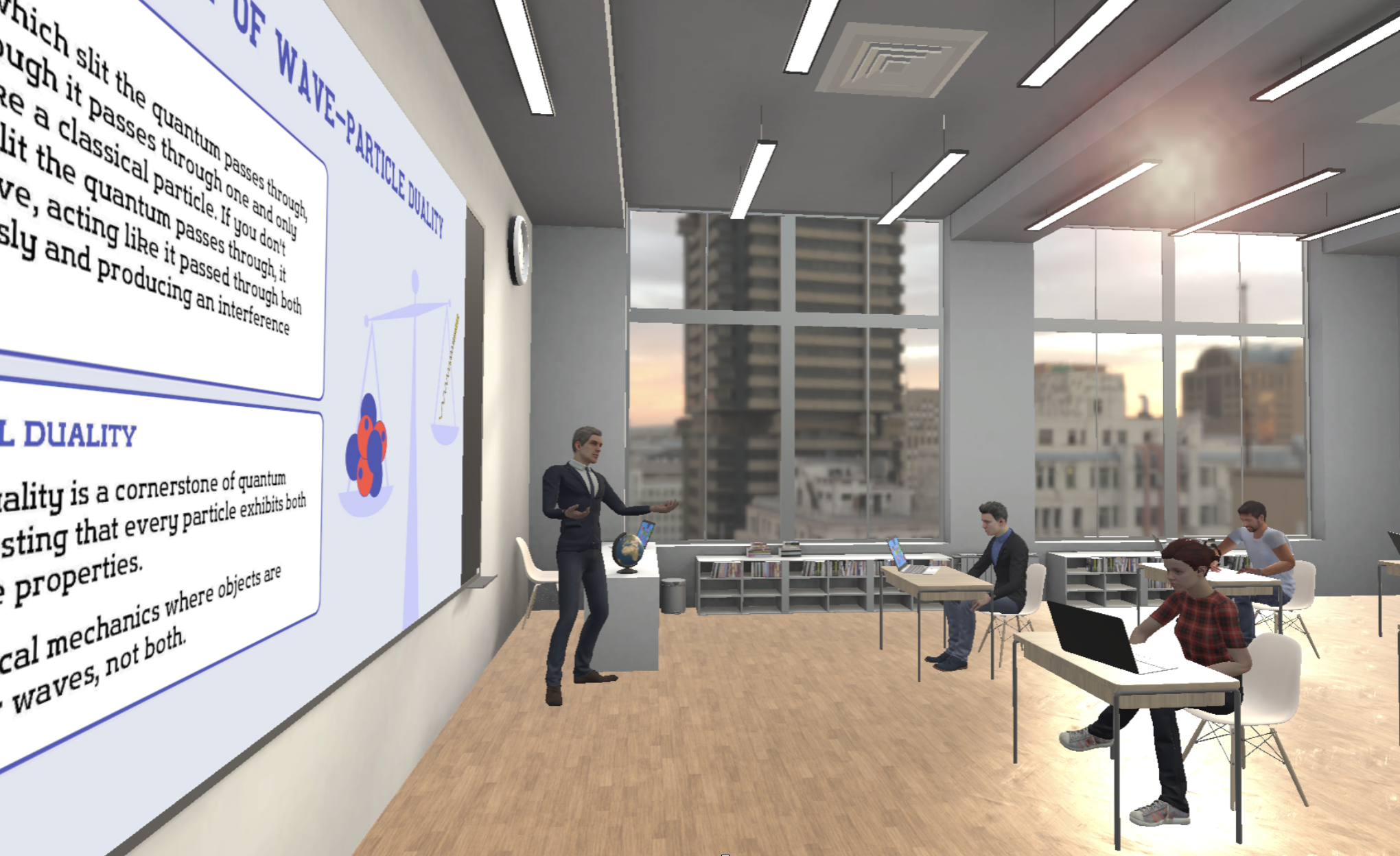}
        
        \Description{Screenshot of a virtual classroom environment, showing a teacher's perspective with virtual students seated in front, digital whiteboards displaying educational content, and an interactive interface.}
        
        \textbf{(a)} View 1 from the virtual classroom.
        \label{fig:vr_fig1}
    \end{minipage}
    \hfill
    \begin{minipage}{0.45\textwidth}
        \centering
        \includegraphics[width=\textwidth]{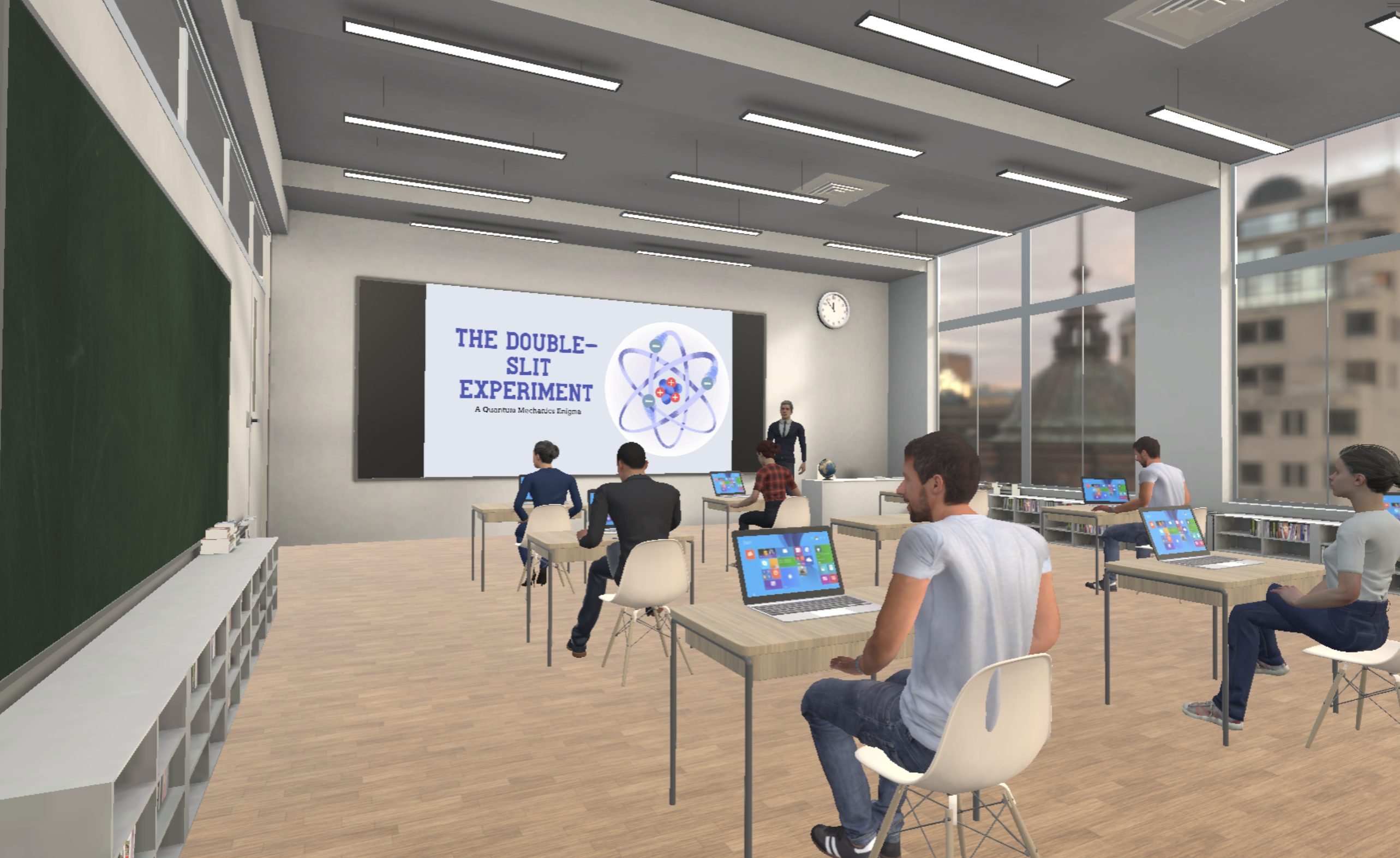}
        
        \Description{An alternate perspective of the virtual classroom, showing a different viewpoint of the interactive environment, including virtual students, instructional materials, and engagement tools.}
        
        \textbf{(b)} View 2 from the virtual classroom.
        \label{fig:vr_fig2}
    \end{minipage}
    \caption{Views from the LLM-driven virtual classroom environment.}
    \label{fig:side_by_side}
\end{figure*}

Education is undergoing a significant digital transformation, accelerated by technological advancements and further driven by the COVID-19 pandemic, which necessitated a shift from in-person to digital learning environments~\cite{zawacki2021current,bygstad2022dual}. Virtual reality (VR) technologies have become increasingly prevalent in this transformation. Advances in VR technology have made head-mounted displays (HMDs) more affordable and accessible, leading to their widespread application across various fields, including healthcare~\cite{javaid2020virtual,pottle2019virtual,kouijzer2023implementation}, entertainment~\cite{chirico2016virtual,ansari2022implementing}, and education~\cite{rojas2023systematic,freina2015literature,bozkir2021exploiting,gao2021digital,huang2023virtual,VR_Handanimation}. In education, VR is revolutionizing traditional teaching methods by transitioning them into a dynamic digital landscape. Institutions like Stanford University have begun conducting entire classes in VR, showcasing VR’s potential to transform conventional teaching~\cite{stanford_vr_class_2021}. Virtual environments enable immersive experiences and enhanced visualizations, which can lead to more effective and engaging learning.

In the digital transformation of education, large language models (LLMs), which are powerful AI systems trained on vast amounts of text data to understand and generate human-like language, are increasingly being applied~\cite{kasneci2023chatgpt,abdalrazaq2023large,yan2024practical,jeon2023large}. In educational settings, integrating LLMs with VR environments enables a more interactive learning experience by allowing students to engage in realistic simulations, ask questions, and receive immediate, contextually accurate responses~\cite{izquierdo2024virtual,liu2024classmeta,kapadia2024evaluation, dataliVR}. These advancements allow for natural conversations and personalized interactions, enhancing the ability to create tailored learning experiences that adapt to individual student needs~\cite{mollick2023using,kasneci2023chatgpt,meyer2024using}. By combining VR with LLMs, educational environments can offer more interactive and flexible learning experiences that accommodate diverse learning styles and preferences, improving student engagement and retention.

Building on advancements in VR and AI-driven systems like LLMs, educational settings have increasingly enabled adaptable and tailored learning experiences that address individual student needs and preferences. Such personalized environments allow students to progress independently and at their own pace, aligning their learning journeys with their unique goals~\cite{pratama2023revolutionizing,shemshack2020systematic,spector2016potential}. However, despite offering substantial flexibility and autonomy, these individualized learning settings often lack the collaborative dynamics characteristic of traditional classroom environments. Peer interactions, particularly question-asking behaviors, significantly enhance the learning experience by capturing students’ attention, fostering deeper engagement, and encouraging cognitive elaboration~\cite{king1989effects,lou2001small,alharbi2022effect}. These interactions function as instructional signals, effectively guiding learners toward critical instructional content, focusing their attention, and enhancing comprehension~\cite{sweller2011cognitive,mautone2001signaling,glynn1979control}. Thus, incorporating peer questions into virtual learning environments can not only direct student attention to crucial concepts but also enrich engagement and foster meaningful interactions within the learning process.

Addressing these challenges and building on technological advancements, our study focuses on integrating LLMs into virtual learning environments to enhance the realism and effectiveness of VR-based education. While previous research has explored the use of VR in education~\cite{bozkir2021exploiting,gao2021digital,lin2024impact,rojas2023systematic,jin2022will}, particularly in simulating real-world environments and improving engagement, the integration of LLMs to create more interactive and personalized experiences has not been widely studied. Few studies~\cite{liu2024classmeta} have examined how LLM-driven interactions within VR classrooms can mirror the dynamics of traditional classrooms, particularly in peer-to-peer or teacher-student exchanges. Our work takes a first step toward addressing this gap by simulating a fully LLM-driven virtual environment (see Figure~\ref{fig:side_by_side}), where an LLM-powered teacher delivers content based on provided slides, complemented by LLM-powered peer interactions. This setup aims to create an immersive learning experience that mirrors real-world educational settings. To evaluate the impact of this integration, we test two conditions: LLM-driven Peer Interaction with Questions and Answers (Peer-QnA), where LLM-driven peers (students) ask questions to the teacher, creating a more interactive classroom environment, and LLM-driven Peer without Question and Answer Interaction (Peer-NoQnA), where LLM-driven peers do not ask questions, leaving the participant as the only entity interacting with the teacher. To understand the effects of these two settings, we evaluated factors including student engagement, cognitive load, and eye-tracking behaviors within the VR.  

The contributions of this work are fivefold: (1) we designed a fully LLM-driven virtual classroom environment and collected data from 19 participants, demonstrating the effectiveness of LLM-driven interactions in immersive educational settings; (2) we evaluated two distinct interaction conditions, Peer-QnA and Peer-NoQnA, and found significant differences in student attention, with the Peer-QnA condition leading to increased attention and engagement with the learning content; (3) we analyzed cognitive load using NASA Task Load Index (NASA-TLX) assessments, supplemented by eye-tracking data, revealing that the Peer-QnA condition resulted in higher cognitive load, as evidenced by increased pupil diameter, which strongly correlated with student attention on crucial content; (4) peer questions led to more targeted attention, resulting in longer mean fixation duration and shorter average saccade amplitude; and (5) in more complex subjects, these changes in cognitive load and visual attention were more noticeable, emphasizing the importance of subject complexity.


\section{Related Work}
\label{sec:related work}
In the evolving educational technology landscape, immersive VR and LLMs have become transformative tools that significantly impact learning environments. In the following subsections, we review the literature in these two areas, highlighting how each contributes to advancements in educational technology. 

\subsection{VR Environments in Education}
Virtual learning spaces are increasingly integrated into educational settings~\cite{radianti2020systematic,rojas2023systematic}, providing valuable insights from both teacher and student perspectives~\cite{papaioannou2023learning,naimi2023teaching}. For teachers, VR allows them to simulate complex, real-world scenarios, allowing them to practice classroom management and lesson delivery in a controlled environment~\cite{peterson2018enhancing,huang2021classroom}. This improves their confidence and teaching strategies before entering a real classroom. From the student perspective, VR enhances engagement by creating immersive, interactive environments that support experiential learning, enabling students to explore concepts in a more hands-on way than traditional methods~\cite{zhang2023investigating,mystakidis2020distance,christopoulos2018increasing}.

In addition to offering general classroom preparation, particularly for teachers in training, VR environments provide a unique opportunity to simulate complex classroom scenarios that can help develop essential teaching and classroom management skills. To this end, Westphal et al.~\cite{westphal2024more} found that student teachers who focused on self-reflection using first-person pronouns in a VR environment experienced higher stress levels, which caused increased stress in subsequent teaching sessions. This highlights the psychological impact of self-focused reflection in VR settings and the importance of preparing teachers to manage stress effectively. Similarly, Huang et al. \cite{huang2021classroom} examined how complex classroom environments, characterized by multiple and overlapping disruptions, impact student teachers’ ability to detect and respond to those disruptions. Their study reveals that higher complexity reduces the likelihood of noticing and effectively addressing disruptions, underscoring the potential of VR to simulate challenging teaching environments that can better prepare teachers for real-world scenarios. Additionally, Huang et al.~\cite{huang2022class} explored the effect of class size on stress levels in pre-service teachers. The authors find that larger class sizes in a VR classroom significantly increase heart rate and perceived stress, indicating that VR effectively simulates classroom management challenges, helping educators develop the skills needed to handle real-world classroom demands.

From the student perspective, VR environments have been shown to enhance engagement, motivation, and overall learning outcomes. Liu et al.~\cite{liu2022effects} demonstrate that primary school students in an immersive VR classroom achieved higher academic success and increased science motivation compared to those in traditional classrooms while also experiencing reduced cognitive load. This suggests that VR can create more engaging and less mentally taxing learning environments for students. Furthermore, Gao et al.~\cite{gao2021digital} and Bozkir et al.~\cite{bozkir2021exploiting} explore how various factors, such as student seating positions, visualization styles, and hand-raising behaviors of virtual peers, impact students’ engagement and attention in a VR classroom. The findings reveal that students seated at the back of the classroom struggle to effectively extract information, while realistic visualization styles of avatars lead to better engagement with lectures. 


Building on previous research that has examined student behaviors and interactions in VR classrooms, Hasenbein et al.~\cite{hasenbein2023investigating} investigate how students interact with social comparison information, particularly in relation to peers' achievement-related behaviors. The authors' findings indicate that students who spend more time observing their peers' achievements tend to have lower self-evaluations, highlighting the psychological effects of peer interactions in virtual settings. Stark et al.~\cite{stark2024using} examine student interactions by using gaze entropy to identify and differentiate classroom discourse events. They are able to predict teacher-led activities and explanations with a high degree of accuracy, demonstrating the potential of gaze entropy as a tool for analyzing classroom participation and engagement in VR settings. While existing studies emphasize the immersive potential of VR in education, there is a research gap regarding fully LLM-driven, dynamic individual learning environments that closely resemble real-world educational experiences. The effects of these AI-driven interactions on student engagement, attention, and learning outcomes are still largely unexplored.

\subsection{Large Language Models in Education}

Recent advancements in LLMs significantly expand their applications across various fields, including healthcare~\cite{thirunavukarasu2023large}, education~\cite{kasneci2023chatgpt}, and beyond~\cite{zhao2023survey,wu2024survey}. In the education domain, these models are now playing a larger role in enhancing both teaching and learning experiences. These models provide personalized learning opportunities, helping students learn independently and adapt to their unique needs, thereby contributing to more equitable education \cite{kasneci2023chatgpt,yan2024practical,park2024promise,han2024teachers}. 

LLMs have demonstrated their versatility and effectiveness across various educational levels, from primary schools to universities. For instance, Yan et al.~\cite{yan2024practical} identify 53 different application scenarios where LLMs are used to automate educational tasks at different levels, including assessment and grading, teaching support, and knowledge representation. This broad applicability highlights the potential of LLMs for innovating traditional educational processes. At the university level, Abd-alrazaq et al.~\cite{abdalrazaq2023large} explore the potential of LLMs in medical education, noting their ability to innovate curriculum design, teaching methodologies, and student assessments. Similarly, in secondary education, Lieb and Goel~\cite{lieb2024student} introduced NewtBot, a personalized tutor chatbot for physics students, which provides positive learning experiences and demonstrates the potential of LLMs as effective virtual tutors. Moreover, Lu and Wang~\cite{lu2024generative} utilize these models to simulate student profiles for evaluating multiple-choice questions. The authors find that the simulated responses are consistent with real student answers, aiding in the refinement of question quality.

LLMs have also been utilized in enhancing interactive learning environments. Liu et al.~\cite{liu2024classmeta} introduced ClassMeta, a GPT-4 driven agent that simulates an active student in a VR classroom. This integration of LLMs with VR significantly expands engagement and learning outcomes, providing a more immersive and effective educational experience. Similarly Izquierdo-Domenech et al.~\cite{izquierdo2024virtual} combined VR and LLMs to create context-aware educational experiences. Participants using this integrated setup achieve significantly better learning outcomes compared to those using traditional methods, highlighting the potential of LLMs to enhance interactive learning.

Despite the promising advancements in integrating LLMs with VR environments~\cite{bozkir2024embedding}, research in this area is still in its early stages. While there have been successful applications of LLMs for tasks like individual tutoring and knowledge representation, there is still much to explore the effects of interacting with AI-powered peers and teachers in immersive settings have not been fully investigated. This highlights the need for further research into how LLMs can enhance not only individual learning experiences but also create more engaging and interactive virtual classrooms. Our study takes an initial step toward addressing these gaps by exploring how LLMs can be integrated into VR to simulate more interactive and realistic classroom dynamics.

\section{Methodology}
\label{sec:method}
As LLM-driven classrooms become more common, it is important to understand how students interact and learn in these new environments. The main purpose of this study is to evaluate student behaviors in fully LLM-driven classroom settings and to analyze the impact of LLM-driven peer questions on engagement, attention, cognitive load, and learning outcomes. Understanding how student attention and cognitive load evolve in these virtual environments can guide the design of more effective and engaging LLM-driven learning spaces. In this section, we provide an overview of the participant details, apparatus, experimental design, procedure, measurement techniques, and data pre-processing steps.

\subsection{Participants}
The study included 19 participants with a mean age of 25.32 and standard deviation  $8.57$, with a gender distribution of 68.42\% male (\(n = 13\)) and 31.58\% female (\(n = 6\)). Educational backgrounds varied, with 63.16\% holding a bachelor’s degree, 26.32\% having completed high school or equivalent, and 10.53\% possessing a master’s degree. Occupationally, 68.42\% were students, while 31.58\%  were recent graduates employed in various industries. Most participants (68.42\%) had prior experience with VR; however, only 5.26\% had used VR in educational settings. Additionally, 94.74\% had interacted with LLMs before, indicating a certain level of familiarity with AI technologies.

\subsection{Apparatus}
The study was conducted using a VR classroom environment designed in Unity3D (see Figures~\ref{fig:side_by_side}). The virtual classroom was equipped with avatars representing a teacher and students, all powered by LLMs to simulate real-time interactions. Specifically, we utilized ``ChatGPT-4o''~\cite{gpt4o} to power the interactions within the classroom. For speech recognition, we employed OpenAI’s Whisper API~\cite{openai_whisper_api} for speech-to-text conversion and Amazon Polly~\cite{amazon_polly} for text-to-speech synthesis. This environment was designed to mirror a realistic classroom, with all avatars equipped with animations to enhance realism. Student avatars featured both speaking and idle animations, while the teacher avatar included additional variations of idle and speaking animations. For question-asking behavior, student avatars raise their hands before speaking, while the teacher avatar uses both a selection gesture and verbal cue when calling on a student. Participants took on the role of a student seated at the center of a $3\times3$ grid, surrounded by eight desks assigned to LLM-driven peers. Participants could ask questions using the HTC Vive controller; pressing the trigger button initiated the speech input system, allowing them to verbally ask their question. The study was conducted with a Varjo XR-3~\cite{varjo_xr3} mixed reality HMD paired with a desktop featuring a 13th Gen Intel Core i7-13700K processor, 32.0 GB of RAM, and an NVIDIA GeForce RTX 4080 GPU. Eye-tracking data was collected using the Varjo XR-3’s built-in eye tracker, operating at the maximum sampling rate of 200Hz. 

\subsection{Experimental Design}
In the study, we evaluated the VR classroom environment by addressing the effectiveness of the virtual setting, which is essential for understanding how well it replicates a real classroom and how immersive and engaging it is for individual participants. Additionally, two separate conditions were tested to examine the effects of interactive dynamics. In the first condition, the participant could address the teacher and ask questions, but the LLM-powered students did not interact. In the second condition, both the participant and the LLM-powered students could address the teacher by asking questions. In both conditions, the teacher presented a set of instructional slides, explaining the content of each one. After the explanation of each slide, a structured opportunity for questions followed. In the Peer-NoQnA condition, the participant could ask a question using a button on the controller, and the teacher responded accordingly. In the Peer-QnA condition, the participant could still ask a question using the same method. Additionally, one or two randomly selected AI student avatars also asked questions after each slide, regardless of whether the participant chose to ask one.

In this study, we employed a within-subjects experimental design, where each participant experienced two interaction conditions within a virtual reality classroom. For each participant, to avoid content repetition, each condition was associated with a different topic, and the order of topic presentation was counterbalanced across participants to mitigate potential order effects. This design ensured that any observed differences in engagement or learning outcomes were attributable to the interaction model rather than the sequence in which the topics were presented. Although we observed variation in participant behavior and outcomes across the two topics, topic complexity was not manipulated as an independent variable. Instead, topic assignment served solely as a counterbalancing mechanism. Nonetheless, the differences observed across topics provide valuable exploratory insights into how content complexity may affect attention, cognitive load, and learning outcomes. 

In the experiment, participants were exposed to four cases involving two topics: the Double-Slit Experiment and the History of Video Games. We choose those topics to allow for an analysis of user behavior in both a technical and a less specialized or non-technical subject, providing deeper insights into how the virtual environment performs across different types of content. Each case varied based on whether questions were asked solely by the user or by the user and AI students. In Case 1, only the participant was allowed to ask questions during the Double-Slit Experiment, whereas both the participant and AI students could ask questions during the History of Video Games. Case 2 reversed this setup, with both the participant and AI students asking questions during the Double-Slit Experiment, followed by only participant questions during the History of Video Games. Case 3 and Case 4 mirrored the structure of Case 1 and Case 2, respectively, but with the order of topics reversed.

In the experiment, we collected eye-tracking data to analyze student behavior in the virtual reality classroom, following the approach used in other studies~\cite{rappa2022use,stark2024towards,hasenbein2022learning,bozkir2021exploiting}. This method provided valuable insights into how participants directed their attention and interacted with various elements of the virtual environment. Additionally, the cognitive load was assessed to measure the mental effort required, consistent with approaches in similar studies~\cite{liu2022effects,albus2021signaling}. The NASA Task Load Index (NASA-TLX), a widely recognized tool for evaluating cognitive load~\cite{criollo2024analysis,bueno2021effects}, was used to gather subjective assessments of participants’ mental demand, effort, and overall workload. These measures offer a comprehensive approach to understanding attention, engagement, and mental effort in the virtual learning environment, validating the use of eye-tracking and cognitive load as key tools for this evaluation. Additionally, we administer pre- and post-questionnaires to gather participants’ feedback on their experience in the virtual classroom environment.

\subsection{Procedure}
Upon arrival, participants were welcomed and asked to complete informed consent forms and a pre-questionnaire with demographic questions. Following this, they were introduced to the first condition of the experiment, conducted in a VR classroom environment. After completing the first condition, participants took the NASA-TLX test to assess the task load. They then proceeded to the second condition, again in the VR classroom, followed by the NASA-TLX test once more. Finally, participants completed a post-questionnaire that gathered general feedback on the overall experience. To ensure participants remained attentive to the lecture content in the virtual classroom, a set of questions related to the presented topic was asked after each condition, assessing their retention and engagement. Each VR session took approximately 15-18 minutes, and the total duration of the experiment was about 1 to 1.5 hours. Participants received compensation for their time and participation.

\subsection{Measurements}
Data collection in this study involved multiple methods to comprehensively assess participant experiences and outcomes.

\subsubsection{Visual Scanpath}  
Eye-tracking data was collected using the Varjo XR-3 headset’s built-in capabilities. From this data, we extract fixation points, where the gaze remains focused for a significant period, and saccades, which are rapid eye movements between fixation points. We analyze key metrics such as total fixation duration, mean fixation duration, saccade amplitude, and saccade velocity. 

The eye-tracking data is crucial for understanding the behaviors of the students in the virtual classroom ~\cite{gao2021digital,bozkir2021exploiting,gao2023exploring,ferdinand2024impact}. Total fixation durations represent the user’s attention to specific content and areas of interest, indicating how long the information is engaged with~\cite{hahn2022eye}. In our experiment, we normalize these durations by dividing them by the total fixation duration for each participant to evaluate attention. The mainboard and teacher were identified as the primary instructional content, and we designated them as the key areas of interest for analyzing how participants focused on the content. Mean fixation duration serves as an indicator of cognitive processing demands~\cite{just1976eye,hahn2022eye}. Higher values generally suggest that deeper cognitive processing is required to process the information being viewed. It is engaged in more complex mental activities, such as understanding, analyzing, or integrating different pieces of information, which also supports meaningful learning~\cite{rayner1998eye,seufert2003supporting}.
Saccade amplitudes provided insight into how broadly or narrowly participants scanned the environment, indicating their visual exploration patterns. Larger saccade amplitudes suggest participants were scanning across a wider area, while smaller amplitudes indicate higher cognitive load with a more concentrated focus on particular content~\cite{chen2011eye}. Saccade velocities indicate how cognitive load and task demands impact attention and engagement. As cognitive load rises, saccade velocity tends to increase. This indicates deliberate focus, where participants take more time to process detailed information or engage more with the content~\cite{gibaldi2021saccade} However, higher average saccade velocities are often associated with increased stress and reduced concentration during cognitive tasks~\cite{behroozi2018dazed,lewandowska2022eye}. In addition to fundamental fixation and saccade metrics, we examined the total fixation duration on key objects, such as the mainboard and teacher, to assess students’ attention. These objects were identified as the primary sources of instructional content, reflecting where the core learning material was delivered.

\subsubsection{Cognitive Load}
Cognitive Load Theory~\cite{sweller2011cognitive,sweller2010element} is a framework for understanding the mental effort involved in learning. Cognitive load is classified into three categories: intrinsic, extraneous, and germane. Intrinsic load is primarily related to the inherent difficulty of the material. Extraneous load arises from unnecessary complexity in the environment. Germane load results from processing information to support understanding~\cite{kirschner2002cognitive,van2010cognitive}. Effective instructional design focuses on minimizing extraneous load while enhancing germane load, enabling learners to concentrate on the essential material without being overwhelmed. In this study, we evaluate the impact of the fully LLM-driven virtual classroom on participants’ cognitive load to understand how interacting with AI-driven peers and teachers influences mental effort. The NASA-TLX was used after each condition to assess participants’ perceived workload. In addition to measuring cognitive load with NASA-TLX, pupil diameter was also utilized to assess the cognitive load experienced by participants during the experiment, serving as an objective indicator of cognitive effort~\cite{beatty1982task,krejtz2018eye,kiefer2016measuring,bozkir2023eye}.

Additionally, we investigate the relationship between cognitive load and visual attention using Pearson correlation and linear regression analyses. The Pearson correlation measured the strength and direction of the association between cognitive load, as assessed by NASA-TLX, and normalized fixation duration on primary instructional elements. Following the correlation analysis, a linear regression was conducted to predict cognitive load based on normalized fixation duration on these key instructional areas. This regression analysis quantified how much variance in cognitive load could be explained by participants’ attention to these primary content areas. 

\subsubsection{Questionnaires}  
We administered several questionnaires in the study, and their details are as follows. 

Pre-Questionnaire includes demographic questions to capture participants' background information, including their age, gender, education level, prior experience with VR and AI technologies, and familiarity with the subject.
Knowledge questionnaire has a multiple-choice test designed to assess their understanding and retention of the content presented during the VR sessions. To investigate the relationship between visual attention and learning outcomes, we conducted a regression analysis.
Post-Questionnaire was administered to gather participants’ overall impressions and feedback on their experience within the virtual classroom environment. This questionnaire was divided into five categories and employed a 5-point Likert scale with response options: ``strongly disagree,'' ``disagree,'' ``neutral,'' ``agree,'' and ``strongly agree.''The questionnaire was structured into five categories. First, we focused on ``Technical Challenges and Audiovisual Quality'', addressing any technical issues participants faced that could have impacted their engagement. Second, we evaluated the Teacher-Student Interaction Quality, focusing on the clarity of the LLM-driven teacher’s content delivery and the effectiveness of its responses during interactions. Third, we examined ``Student Participation and Peer Influence'', particularly how the question-and-answer dynamics between peers and the teacher affected participants’ attention, engagement, and overall learning process. Fourth, we assessed ``Assessment Quality and Relevance'', collecting participants' feedback on the appropriateness and difficulty of the test questions used to evaluate their understanding of the lecture content. Finally, we collected feedback on Overall Experience and Satisfaction, reflecting participants’ general impressions of the LLM-driven VR classroom environment.

\subsection{Data Processing}

Eye-tracking data was processed using the Identification by Velocity Threshold (I-VT) algorithm~\cite{salvucci2000identifying,kasneci2024introduction} to identify fixations and saccades. The I-VT algorithm classifies eye movements by measuring gaze velocity, with slower movements being categorized as fixations and faster movements as saccades. We also incorporated the head movements in the fixation detection, as fixations were only counted when both eye and head movements were stable. The specific criteria used for detecting fixations and saccades, including velocity and duration thresholds~\cite{gao2021digital,agtzidis2019360}, are provided in Table~\ref{table:criteria}. For the pupil diameter data, we applied a Savitzky-Golay filter~\cite{savitzky1964smoothing} to smooth the data and remove noise. Following this, divisive baseline correction with a baseline duration of 1 second~\cite{mathot2018safe} was used to normalize the readings, ensuring a more precise analysis of cognitive load and engagement, as seen in similar studies~\cite{gao2021digital,bozkir2019person,bozkir2019assessment}. Similarly, to analyze the total fixation duration on key objects (mainboard and teacher), we normalized this metric by dividing it by the overall fixation duration for each participant.

\begin{table}[h]
\small 
\centering
\caption{Criteria for Fixation and Saccade Detection.}
\begin{tabular}{@{}p{0.2\columnwidth} p{0.3\columnwidth} p{0.4\columnwidth}@{}}
\toprule
\textbf{Event} & \textbf{Velocity ($v$)} & \textbf{Duration ($\Delta$)} \\
\midrule
Fixation & 
\begin{tabular}[t]{@{}l@{}} 
$v_{head} < 7^\circ/s$ \\ 
$v_{gaze} < 30^\circ/s$ 
\end{tabular} 
& 
\begin{tabular}[t]{@{}l@{}} 
$ \Delta_{fixation} > 100 \ ms $ \\ 
$\Delta_{fixation} < 500 \ ms$ 
\end{tabular} \\
\addlinespace
Saccade & 
$v_{gaze} > 40^\circ/s$ & 
\begin{tabular}[t]{@{}l@{}} 
$\Delta_{saccade}>20 \ ms $ \\ 
$\Delta_{saccade} < 100 \ ms$ 
\end{tabular} \\
\bottomrule
\end{tabular}
\label{table:criteria}
\end{table}

\subsection{Analysis}
We conducted a separate analysis of cognitive load, visual scanpath, and learning outcomes for each topic, the Double-Slit Experiment and the History of Video Games. For both topics, we compared the Peer-QnA and Peer-NoQnA conditions in terms of cognitive load, pupil diameters, fixations, saccades, and learning outcomes. We conducted an independent t-test for normally distributed samples. For distributions that did not conform to a normal distribution, we used the non-parametric Wilcoxon signed-rank test. Normality was evaluated using the Shapiro-Wilk test for sample sizes under 2000~\cite{razali2011power,royston1982extension} and the Kolmogorov-Smirnov test~\cite{lilliefors1967kolmogorov} for larger samples. In all analyses, a significance level of \( \alpha = 0.05 \) was applied to determine statistical significance.

To analyze the questionnaire data, we calculated the mean and standard deviation for each question using the Likert scale, where ``strongly agree'' corresponds to 5 and ``strongly disagree'' corresponds to 1~\cite{joshi2015likert}. A mean score above 3 indicates general agreement or a positive response, while a mean below 3 reflects disagreement or negative feedback. For reverse-worded questions, we adjusted the scoring by inverting the response values during analysis to maintain consistency. Additionally, we calculated internal consistency for each category using Cronbach’s alpha~\cite{tavakol2011making}. This method evaluates how well the items within each category measure the same construct. A Cronbach’s alpha value above 0.8 is considered high reliability, between 0.6 and 0.8 indicates moderate reliability, and below 0.6 suggests low reliability~\cite{pallant2020spss}. A high alpha value indicates strong internal consistency, meaning that the items within each category are reliably measuring the intended construct. We used Cronbach's alpha as a robust reliability indicator, although it can underestimate reliability when applied to a small number of items~\cite{chakrabartty2013best,malkewitz2023estimating, tavakol2011making}.

\section{Results}
\label{sec:results}
We presented the results for each topic's cognitive load analysis, visual-scanpath analysis, and learning outcomes separately. Following this, we provided the findings from a general questionnaire, where user feedback on their overall experience was collected.

\subsection{Topic 1: Double-Slit Experiment}

\subsubsection{Cognitive Load Analysis}
\begin{figure*}[ht]
    \centering
    \begin{minipage}{0.32\textwidth}
        \centering
        \includegraphics[width=0.95\textwidth]{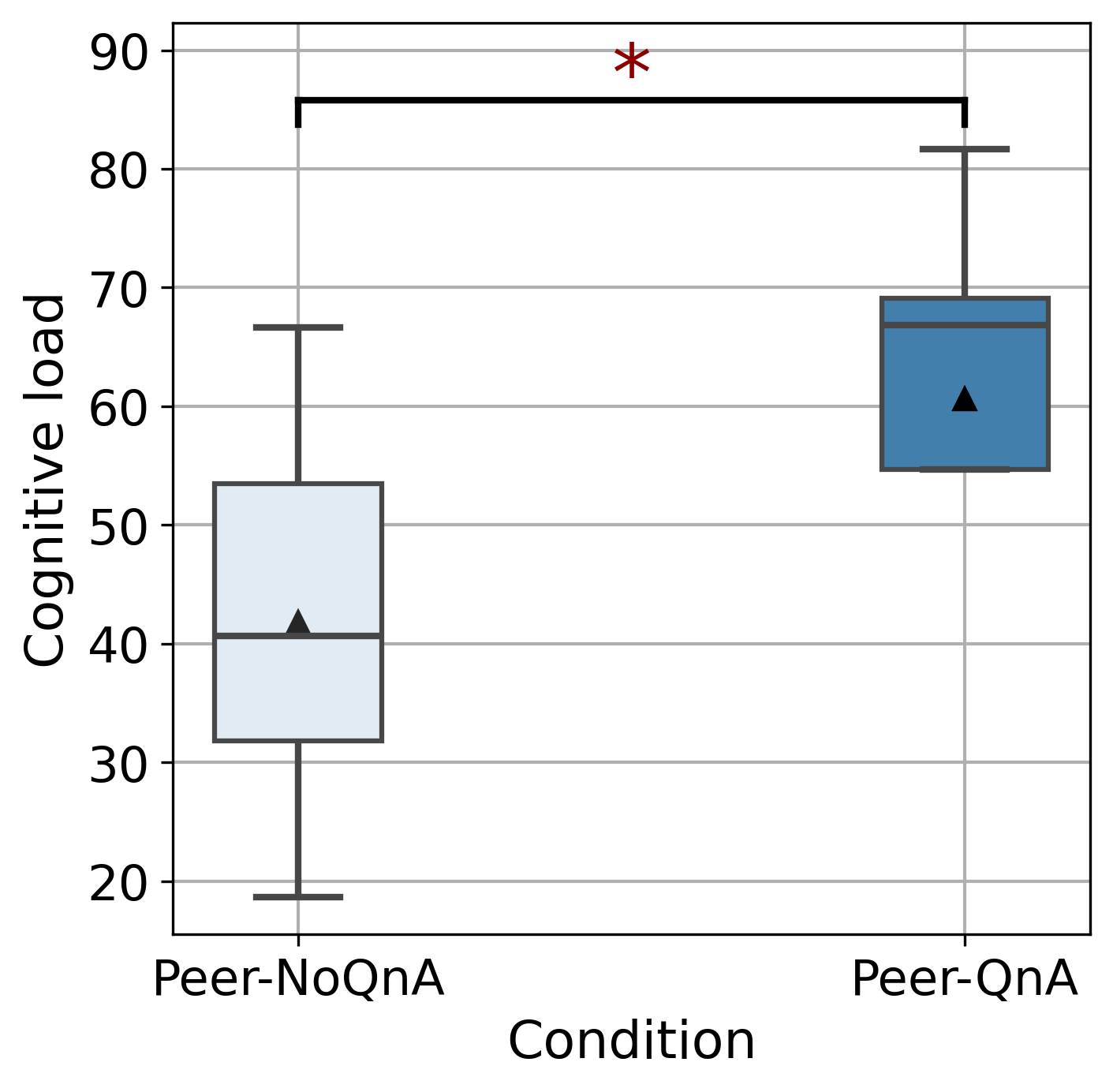}
        \Description{Box plot comparing cognitive load scores for Peer-NoQnA and Peer-QnA conditions in the Double-Slit Experiment. Peer-QnA shows higher cognitive load.}
        \textbf{(a)} Cognitive load scores.
        \label{fig:cognitive_load}
    \end{minipage}
    \hfill
    \begin{minipage}{0.32\textwidth}
        \centering
        \includegraphics[width=0.95\textwidth]{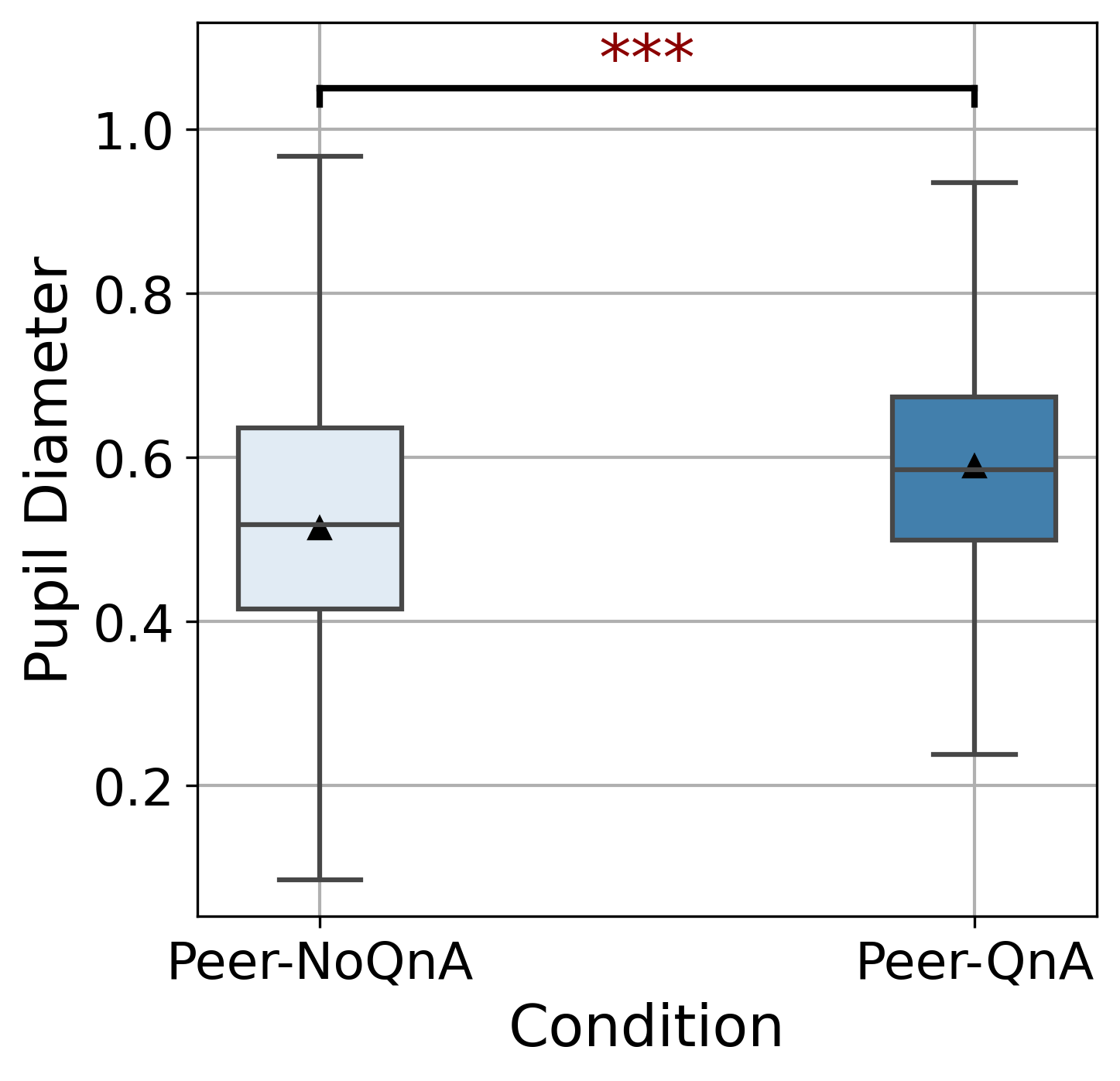}
        \Description{Box plot showing mean pupil diameter distributions under Peer-NoQnA and Peer-QnA conditions. Peer-QnA shows significantly higher values.}
        \textbf{(b)} Mean pupil diameters.
        \label{fig:pupil_diameter}
    \end{minipage}
    \hfill
    \begin{minipage}{0.32\textwidth}
        \centering
        \includegraphics[width=0.95\textwidth]{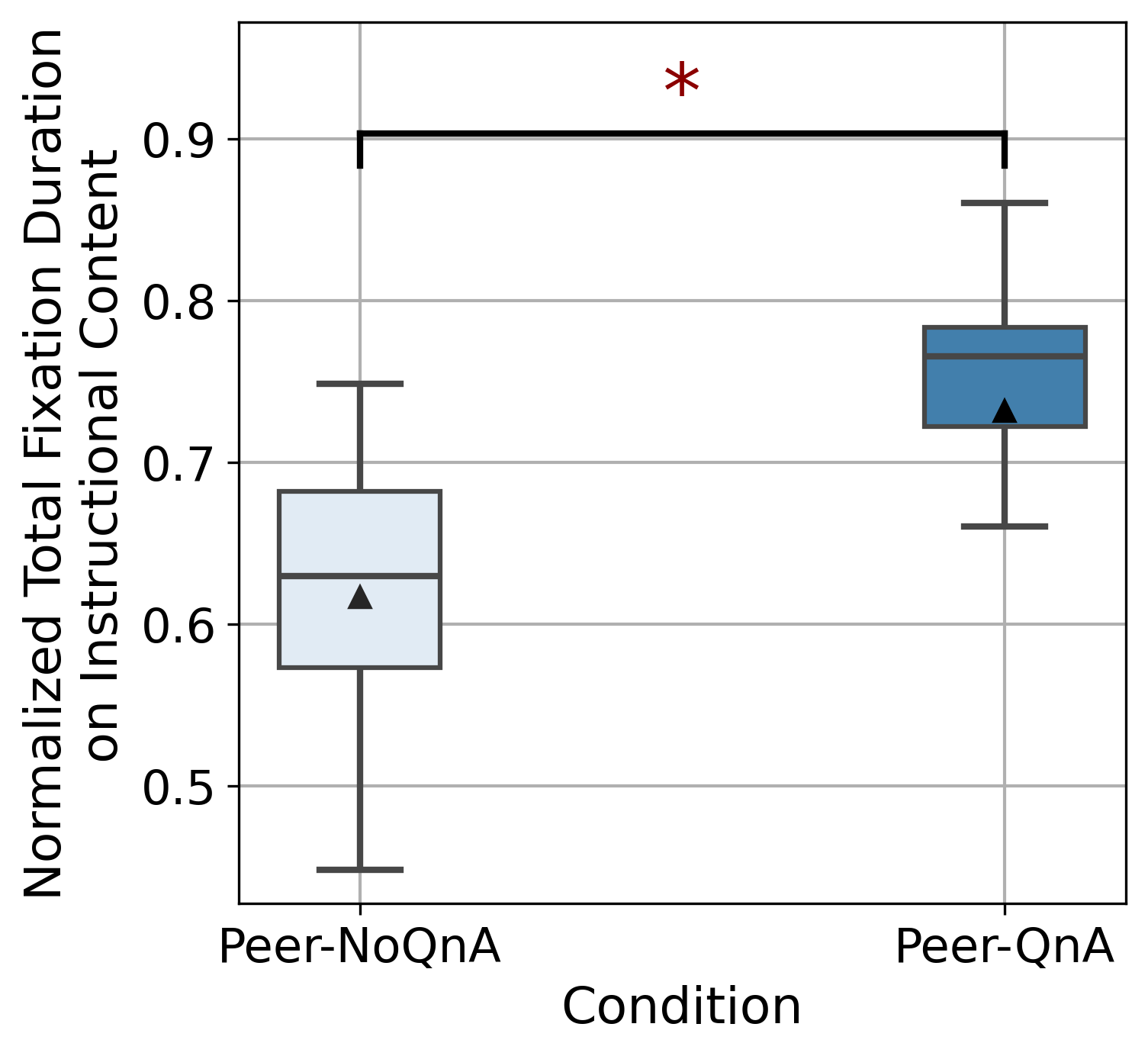}
        \Description{Box plot showing adjusted teacher mainboard duration percentage for the double slit experiment.}
        \textbf{(c)} Fixation duration on primary instructional content.
        \label{fig:violin_plot_teacher_mainboard_double_slit}
    \end{minipage}
    \caption{Results for the Double-Slit Experiment across Peer-QnA and Peer-NoQnA conditions.}
    \label{fig:subfigures_double_slit}
\end{figure*}

In the Double-Slit Experiment, the interactivity of LLM-driven peers significantly impacted cognitive load, as assessed by the NASA-TLX. We observed significantly higher cognitive load scores in the Peer-QnA condition (\(M = 60.50,\) \( SD = 18.16\)) compared to the Peer-NoQnA (\(M = 41.73,\) \( SD = 15.36\)). This difference was statistically significant, with a p-value of \emph{p} = .026 (\(p < .05\)), as given in Figure~\ref{fig:subfigures_double_slit} (a). This finding is further supported by pupil diameters, with significantly higher mean pupil diameters in the Peer-QnA condition (\(M = .59, SD = .11\)) compared to the Peer-NoQnA condition (\(M = .51, SD = .19\)), indicating a significant difference (\emph{p} < .001), as shown in Figure~\ref{fig:subfigures_double_slit} (b).

In our regression analysis, we identified a significant relationship between cognitive load and the normalized total fixation duration on the primary instructional content. The Pearson correlation coefficient was \(r(18) = 0.60\), \(p = .0067\), indicating a strong positive correlation between these variables. As the total fixation duration on the primary instructional content increased, cognitive load also increased. The regression model explained a significant portion of the variance in cognitive load, with \(R^2 = .36\), adjusted \(R^2 = .32\), \(F(1, 18) = 9.53\), \(p = .007\), indicating that 32.2\% of the variance in cognitive load could be attributed to participants' fixation duration on the main instructional content.

\subsubsection{Visual Scanpath Analysis}
A significant difference in fixation durations is observed between the Peer-QnA and Peer-NoQnA groups, with the Peer-QnA condition showing slightly higher means. The mean fixation duration for the Peer-QnA condition is \(M = 233ms,\)  \(SD = 103ms\), compared to \(M = 228ms, SD = 101ms\) for the Peer-NoQnA condition, indicating a statistically significant difference (\emph{p} < .001).

A significant difference is found between the Peer-QnA and Peer-NoQnA conditions in saccade amplitudes. The mean saccade amplitude for the Peer-QnA condition is lower, \(M = 135.99^\circ, SD = 68.74^\circ\), compared to \(M = 137.87^\circ, SD = 69.89^\circ\) for the Peer-NoQnA condition, with a statistically significant difference (\emph{p} = .015, \(p < .05\)). There are no significant differences in saccade velocities between the Peer-QnA and Peer-NoQnA groups. The mean saccade velocity for the Peer-QnA condition is \(M = 127.22^\circ/s, SD = 68.70^\circ/s\), while for the Peer-NoQnA condition, it is \(M = 128.29^\circ/s, SD = 69.34^\circ/s\), with no statistically significant difference.

Additionally, we observed that participants gazed significantly more at the primary instructional content. The normalized total fixation duration is significantly higher in the Peer-QnA condition (\(M = 0.72, SD = 0.11\)) compared to the Peer-NoQnA condition (\(M = 0.60, SD = 0.08\)). The t-test results confirmed a significant difference between these two conditions in the Double-Slit Experiment (\emph{p} = .02, \(p < .05\)), as shown in Figure~\ref{fig:subfigures_double_slit} (c).

\subsubsection{Learning Outcome}
The results of the knowledge questionnaire indicate that in the Peer-QnA condition, the scores were higher compared to the Peer-NoQnA condition, although the difference was not statistically significant. For the Double-Slit Experiment, the mean score for the Peer-QnA condition was \(M = 7.50, SD = 2.07\), while the mean score for the Peer-NoQnA condition was \(M = 6.82, SD = 1.60\) as shown in Figure~\ref{fig:combined_mc_score_video_games_double_slit} (a).

\begin{figure}[ht]
    \centering
    \begin{minipage}{0.46\textwidth}
        \centering
        \includegraphics[width=0.85\textwidth]{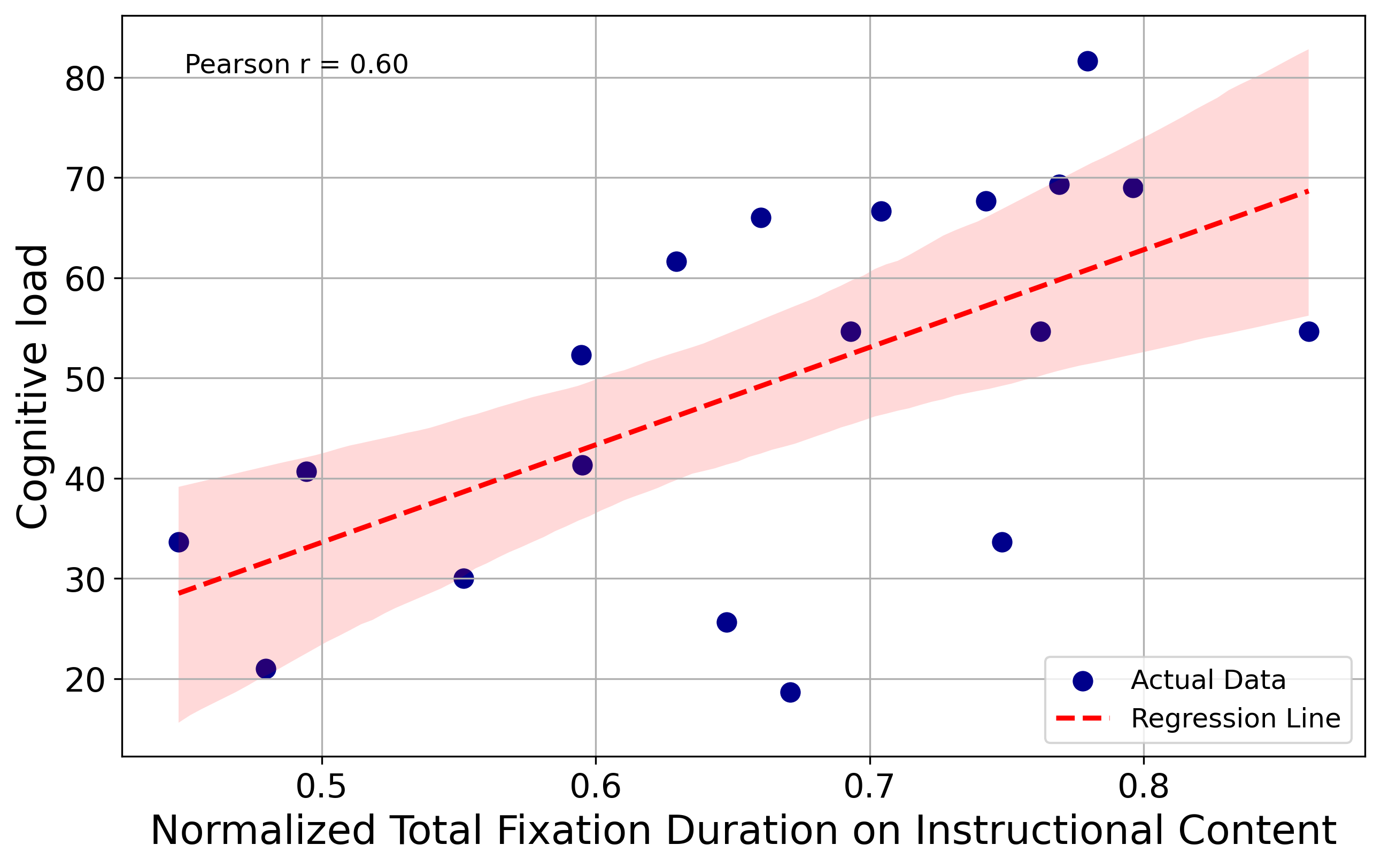}
        
        \Description{Scatter plot showing the relationship between cognitive load and normalized total fixation duration on primary instructional content in the Double-Slit Experiment. A linear regression line is fitted to the data points.}
        
        \textbf{(a)} Cognitive Load vs Normalized Total Fixation Duration on Primary Instructional Content.
        \label{fig:linear_regression_double_slit_teacher_mainboard_duration_percentage}
    \end{minipage}
    \hfill
    \begin{minipage}{0.46\textwidth}
        \centering
        \includegraphics[width=0.85\textwidth]{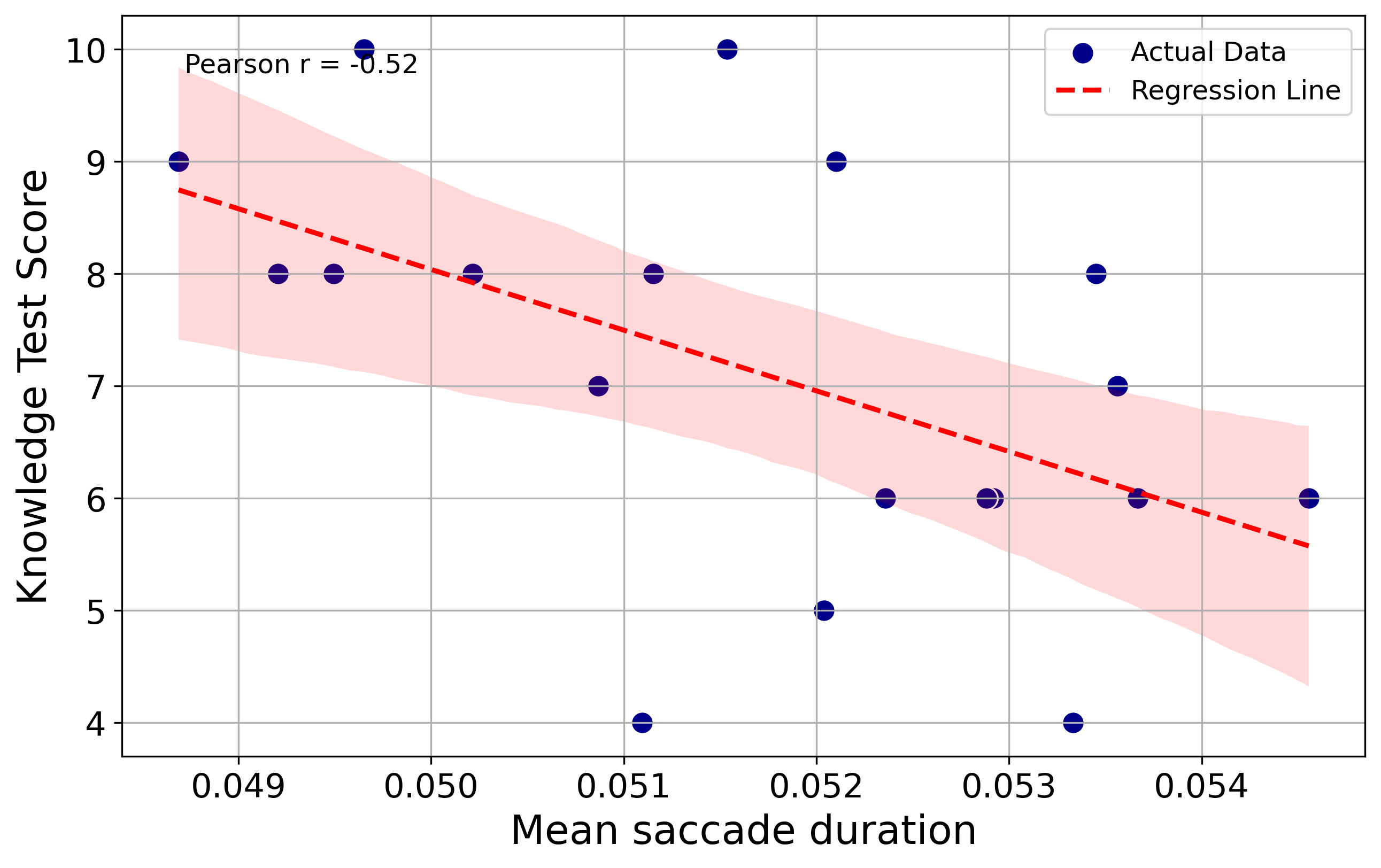}
        
        \Description{Scatter plot showing the relationship between knowledge questionnaire scores and mean saccade duration in the Double-Slit Experiment. A linear regression line is fitted to illustrate the trend.}
        
        \textbf{(b)} Knowledge Questionnaire Scores vs Mean Saccade Duration.
        \label{fig:linear_regression_double_slit_mean_saccade_duration_vs_mc_score}
    \end{minipage}
    \caption{Linear regression results for the Double-Slit Experiment.}
    \label{fig:combined_linear_regression_double_slit}
\end{figure}

Additionally, we applied regression analysis to identify the relationship between visual scanpath metrics and the knowledge questionnaire scores for the Double-Slit Experiment. The analysis revealed a significant negative correlation between mean saccade duration and knowledge scores, with Pearson correlation \(r(17) = -.52\), \(p = .024\), indicating that shorter saccade durations were associated with higher questionnaire scores, as shown in Figure~\ref{fig:combined_linear_regression_double_slit} (b). Linear regression analysis further demonstrated that mean saccade duration explained a significant portion of the variance in the questionnaire scores, \(R^2 = .27\), adjusted \(R^2 = .22\), \(F(1, 17) = 6.19\), \(p = .024\).
When we considered only the mean saccade duration while participants focused on the primary instructional content, the relationship became more pronounced. The Pearson correlation was \(r(18) = -.61\), \(p = .006\), indicating a stronger negative correlation. Mean saccade duration explained 33.6\% of the variance in knowledge questionnaire scores, with \(R^2 = .37\), adjusted \(R^2 = .34\), and the regression model showing statistical significance, \(F(1, 18) = 10.12\), \(p = .005\).

\subsection{Topic 2: History of Video Games}

\subsubsection{Cognitive Load Analysis}
In the History of Video Games, cognitive load metrics were generally lower compared to the Double-Slit Experiment. When comparing the Peer-QnA and Peer-NoQnA conditions, no significant differences in cognitive load were found. The mean cognitive load for the Peer-QnA condition was \(M = 43.58, SD = 15.38\), and for the Peer-NoQnA condition, it was \(M = 42.57, SD = 17.04\). Similarly, normalized pupil diameters showed no significant variation between the two conditions, with \(M = 0.64, SD = 0.11\) for Peer-QnA and \(M = 0.62, SD = 0.11\). These results indicate that, within the context of the video games topic, the level of interactivity had no statistically significant effect on cognitive load or pupil responses. Similarly, correlation and regression analyses revealed no significant relationship between total fixation duration on the primary instructional content and cognitive load.

\subsubsection{Visual Scanpath Analysis}
Similar to the Double-Slit Experiment, a significant difference in fixation durations was found between the two conditions. The mean fixation duration for the Peer-QnA condition was slightly higher (\(M = 237ms, SD = 101ms\)) compared to the Peer-NoQnA condition (\(M = 235ms, SD = 103ms\)), with the difference being statistically significant, \emph{p} = .017 (\(p < .05\)). Saccade amplitudes were similar across both conditions. However, there was a significant difference in saccade mean velocities. The mean velocity for the Peer-QnA condition was slightly higher (\(M = 1.28^\circ/s, SD = 0.71^\circ/s\)) compared to the Peer-NoQnA (\(M = 1.26^\circ/s, SD = 0.67^\circ/s\)), with a statistically significant difference, \emph{p} = .040 (\(p < .05\)). This suggests that participants in the Peer-QnA condition exhibited more rapid saccadic movements compared to those in the Peer-NoQnA condition. We did not observe any significant difference in the normalized total fixation duration on the primary instructional content.

\subsubsection{Learning Outcome}
\begin{figure}[ht]
    \centering
    \begin{minipage}{0.45\textwidth}
        \centering
        \includegraphics[width=0.55\textwidth]{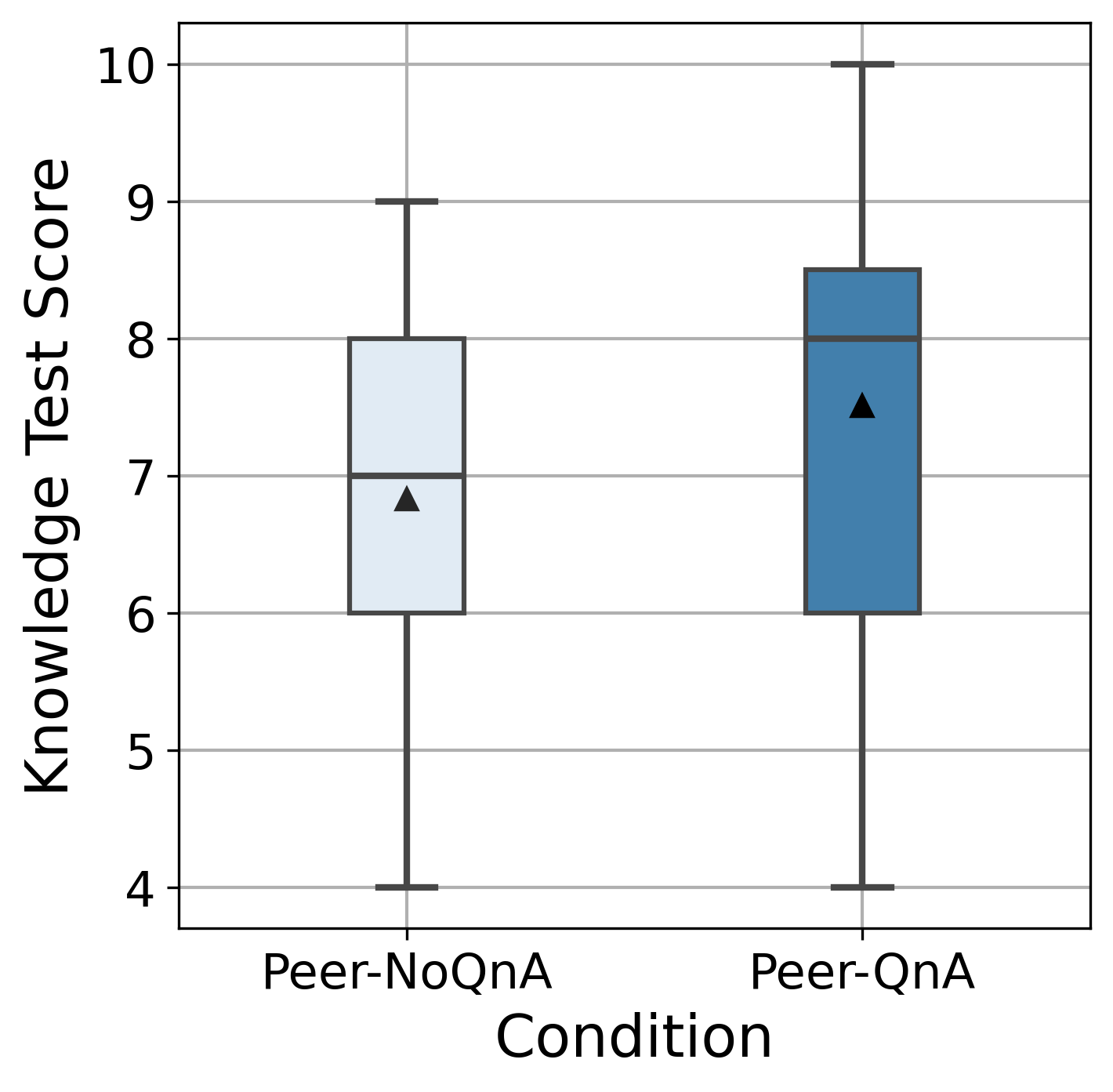}
        
        \Description{Box plot displaying knowledge questionnaire scores for the Double-Slit Experiment under Peer-QnA and Peer-NoQnA conditions. The plot illustrates the distribution of scores, including median and interquartile ranges.}
        
        \textbf{(a)} Knowledge questionnaire scores for Double-Slit Experiment.
        \label{fig:violin_plot_mc_score_double_slit}
    \end{minipage}
    \hfill
    \begin{minipage}{0.45\textwidth}
        \centering
        \includegraphics[width=0.6\textwidth]{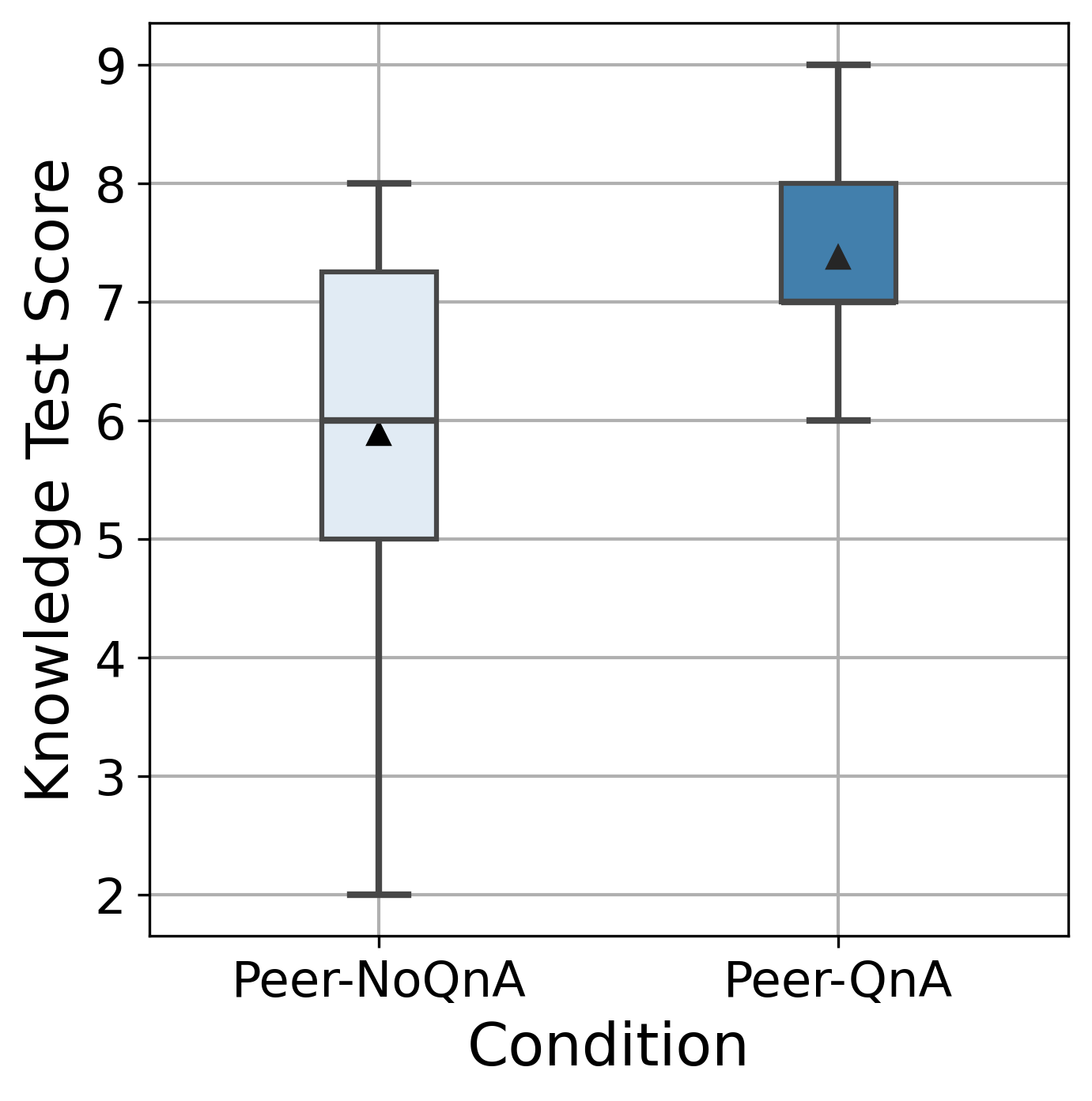}
        
        \Description{Box plot displaying knowledge questionnaire scores for video games under Peer-QnA and Peer-NoQnA conditions. The plot presents the score distribution, highlighting the median and interquartile ranges.}
        
        \textbf{(b)} Knowledge questionnaire scores for History of Video Games.
        \label{fig:violin_plot_mc_score_video_games}
    \end{minipage}
    \caption[Knowledge questionnaire scores comparison]{Knowledge questionnaire scores for Peer-QnA and Peer-NoQnA conditions.}
    \label{fig:combined_mc_score_video_games_double_slit}
\end{figure}

A difference in knowledge questionnaire scores is observed between the conditions. The mean score for the Peer-QnA condition is \(M = 7.36, SD = 1.21\), while the Peer-NoQnA condition has a mean score of \(M = 5.88, SD = 1.96\). Although the difference is not statistically significant, it approached the threshold (\emph{p} = .056), indicating a potential trend towards improved performance in the Peer-QnA condition, as shown in Figure~\ref{fig:combined_mc_score_video_games_double_slit} (b).
In the regression analysis, no significant relationship was found between visual scanpath metrics and knowledge questionnaire scores.

\subsection{General Analysis}
The results of the questionnaire are summarized in Table \ref{table:vr_classroom_stats}, which includes the mean and standard deviation for each question. For each category, Cronbach’s alpha has been calculated to assess the internal consistency of the items, and the corresponding reliability level is provided to indicate the strength of this consistency. Additionally, questions that are reverse-worded are marked with ``(R)''.
\begin{table*}[t]
\small 
\centering
\caption{The means and standard deviations of each question, along with reliability statistics for each category.}
\label{table:vr_classroom_stats}
\begin{tabular}{p{0.75\linewidth} p{0.08\linewidth} p{0.08\linewidth}}
        \toprule
        \textbf{Items} & \textbf{ M} & \textbf{S.D.} \\
        \midrule
        \textbf{Technical Challenges and Audiovisual Quality} (Cronbach's $\alpha$ = .660, Acceptable)\\
        I experienced the latency, and it was disturbing. (R) & 4.4737 & 0.6967 \\
        I experienced technical issues during the session. (R) & 3.7368 & 1.3267 \\
        The audio quality was clear, allowing me to understand the teacher and other students. & 4.3684 & 0.7609 \\
        \midrule
        \textbf{Interaction Quality} (Cronbach's $\alpha$ = .875, High) \\
        The teacher’s slide presentations and explanations were understandable and effective. & 4.0526 & 0.7799 \\
        The content of the teacher’s responses was satisfactory. & 3.2632 & 1.1945 \\
        The responsiveness of the teacher was satisfactory. & 3.5263 & 1.0203 \\
        The teacher’s responses to my questions were adequate and helpful. & 3.5789 & 1.1698 \\
        The teacher made clear explanations. & 3.3684 & 1.1648 \\
        Students’ questions were realistic. & 3.8421 & 0.8342 \\
        The length of the questions was appropriate. & 3.6316 & 1.0651 \\
        Interaction between the teacher and students seemed natural and fluid. & 3.6316 & 1.0116 \\
        \midrule
        \textbf{Student Participation and Peer Influence} (Cronbach's $\alpha$ = .708, Moderate) \\
        The peer interactions make the environment more engaging. & 3.9474 & 0.8481 \\
        The presence of active students in the VR environment enhanced my learning experience. & 4.0000 & 0.8165 \\
        Other students' questions added value to my learning experience. & 3.8947 & 0.9366 \\
        Seeing other students ask questions encouraged me to ask questions as well. & 3.5789 & 0.9612 \\
        Seeing other students ask questions helped maintain my focus on the subject matter. & 4.0000 & 0.6667 \\
        The questions asked by other students in the VR classroom were distracting and made it difficult to maintain focus. (R) & 4.2105 & 0.7873 \\
        \midrule
        \textbf{Assessment Quality and Relevance} (Cronbach's $\alpha$ = .826, High) \\
        Multiple-choice questions regarding the lecture content were comprehensive and relevant. & 4.2632 & 0.6534 \\
        The level of difficulty of the multiple-choice questions was appropriate for my level of understanding. & 3.9474 & 0.8481 \\
        \midrule
        \textbf{Overall Experience and Satisfaction} (Cronbach's $\alpha$ = .794, Moderate) \\
        I felt comfortable interacting in the VR environment. & 3.5789 & 1.1213 \\
        I believe the immersive nature of VR classrooms enhances the learning experience. & 3.9474 & 0.8481 \\
        The VR environment made the subject matter more interesting. & 3.3684 & 1.0116 \\
        VR classroom enhanced my understanding of the material. & 3.2632 & 0.6534 \\
        Using VR technology/experiments changed my perspective on virtual learning positively. & 4.1053 & 0.7375 \\
        I would recommend VR classroom experiences to others. & 4.1579 & 0.7647 \\
        Overall, I was satisfied with my VR classroom experience. & 4.1053 & 0.5671 \\
        \bottomrule
\end{tabular}
\end{table*}

\section{Discussion}

This section discusses the impact of LLM-driven peer interactions on cognitive load, attention, and learning outcomes. We explore how peer-driven questions and content complexity influence these factors on attention, engagement, cognitive load, and learning outcomes. Then, user feedback on the LLM-driven VR classroom environment provides insights into the user experience and design improvements.

\subsection{Cognitive Load, Attention, and Peer \\ Interactions}

The results from the Double-Slit Experiment indicate that the Peer-QnA condition significantly increases cognitive load compared to the Peer-NoQnA condition, as reflected in both NASA-TLX scores and pupil diameter. This increase in cognitive load is attributed to participants’ directed attention toward the primary instructional content, supported by the positive correlation between cognitive load and the normalized total fixation duration on key instructional elements. Peer questions effectively direct participants’ attention to the lecture material, increasing the cognitive effort required as learners engage more deeply with the primary instructional content. These questions do not directly cause extraneous cognitive load; rather, the observed increase in cognitive load is primarily related to the enhanced focus on the lecture material. Furthermore, no significant increase in cognitive load was observed in the History of Video Games topic, also suggesting that peer questions did not introduce extraneous load. This indicates that peer questions do not inherently increase cognitive load; instead, the increase depends on the directed attention to the main lecture content and intrinsic complexity of the material.

In addition to enhancing attention to the main content in the Double-Slit Experiment, the peer questions in the Peer-QnA condition help direct participants to key points within the lesson. This is evidenced by longer mean fixation durations and shorter saccade amplitudes, indicating that participants’ attention is not only more focused but also more precisely targeted. Longer mean fixation durations suggest that participants spend more time processing specific elements of the material, allowing for deeper cognitive engagement~\cite{poole2006eye,hahn2022eye}. Shorter saccade amplitudes, on the other hand, reflect more localized and deliberate eye movements, with participants scanning less broadly across the visual field and honing in on relevant instructional elements~\cite{chen2011eye}. These shorter saccade amplitudes suggest a more efficient and concentrated visual processing strategy, where attention is focused on specific, relevant information without being distracted by peripheral content. Together, these visual attention patterns—longer fixation durations and shorter saccades—indicate better performance~\cite{chen2014eye} and also imply that peer questions acted as signals, guiding participants to focus on critical aspects of the lesson. This aligns with signaling theory~\cite{glynn1979control,loman1983signaling,mautone2001signaling}, which proposes that cues in the learning environment, such as peer questions, can direct learners’ attention to the most important information, thereby enhancing the learning process. In this context, the LLM-driven peer questions in the Double-Slit Experiment function as effective signals, guiding participants to identify and concentrate on the most critical parts of the lesson. These questions contribute to a more focused and targeted learning experience, particularly in the more complex instructional environment. 

Another important point is that the results from the History of Video Games topic show no significant differences in cognitive load, pupil diameter, or total fixation duration between the Peer-QnA and Peer-NoQnA conditions. This highlights the influence of content complexity on attention and cognitive load. The History of Video Games, being less technically demanding, likely does not require the same level of cognitive effort as the Double-Slit Experiment. As a result, the LLM-driven peer questions do not significantly impact attention on the primary content, but there was a noticeable trend toward improved learning outcomes in the Peer-QnA condition. This trend, though not statistically significant, suggests that peer interactions might still support learning even in less complex topics by fostering engagement and verbal processing rather than through increased cognitive load.

These findings have important implications for the design of LLM-driven virtual learning environments. In more complex subjects, like the Double-Slit Experiment, LLM-driven peer interactions can effectively direct attention toward key instructional elements and increase cognitive engagement. In less complex subjects, such as the History of Video Games, peer interactions may not significantly impact visual attention but can still provide educational benefits by promoting engagement and verbal processing, enhancing learning outcomes without introducing extraneous cognitive load.

These findings suggest that the effects of peer interactions may be influenced by content complexity. Although topic complexity was not manipulated as an independent variable, exploratory patterns indicate that peer questions in the more demanding topic, the Double-Slit Experiment, were associated with increased cognitive load, more focused visual attention, and improved learning outcomes. In contrast, for the less complex topic, the History of Video Games, peer interactions had no significant effect on cognitive load or attention metrics, though a trend toward improved learning outcomes was observed. These exploratory insights highlight the need for future research to explicitly manipulate topic complexity and examine interaction effects more rigorously.

The observed negative correlation between mean saccade duration and knowledge questionnaire scores indicates that shorter saccade durations are associated with higher learning outcomes. This suggests that participants exhibiting more rapid eye movements between fixations processed the instructional content more effectively. Shorter saccades typically reflect more focused and efficient visual scanning, facilitating quicker identification and engagement with key information. As a result, this visual processing efficiency likely contributed to improved retention and comprehension of the material. These findings emphasize the critical role of visual attention dynamics in enhancing learning outcomes within virtual learning environments.

\subsection{User Feedback and Design Implications}
The post-experiment questionnaire offers key insights into participants’ experiences within the fully LLM-driven virtual classroom, revealing both positive aspects and areas for improvement. Participants generally rated the technical aspects of the VR environment positively. However, ``I experienced technical issues during the session. (R)'' ($M = 3.74$) is reported as relatively low by some participants, indicating some issues. These were primarily related to the speech-to-text functionality. This emphasizes the importance of system robustness, as those problems could impact the practical use of the technology in real-world educational settings. The interaction quality between LLM-driven peers and the teacher is mostly rated positively. However, some questions, such as ``The content of the teacher’s responses was satisfactory'' ($M = 3.26$), score lower, suggesting room for improvement in response quality. The quality of responses could be enhanced by employing prompting strategies that provide more comprehensive content and by using advanced techniques such as retrieval-augmented generation (RAG), which can deliver more accurate information and reduce hallucinations~\cite{huang2023survey,lewis2020retrieval}. Despite this, the overall interaction quality received positive feedback, with positive scores across items like ``The teacher made clear explanations'' and ``The teacher’s responses to my questions were adequate and helpful.'' In general, Peer interactions were well-received, with most participants finding them engaging rather than distracting. For instance, ``Peer questions encouraged me to ask questions as well'' was rated positively ($M = 3.57$), indicating that peer involvement promotes engagement and participation. Participants also agreed that peer questions enhanced their focus and learning experience, as reflected in statements like ``The presence of active students in the VR environment enhanced my learning experience'' ($M = 4.0$). These findings suggest that incorporating peer interactions makes the learning environment more dynamic and engaging. However, there is room for improvement in exploring interaction strategies, such as peer-to-peer interaction, which could offer additional benefits. Furthermore, the effectiveness of these interactions may also depend on the individual learning styles of the participants~\cite{akkoyunlu2008study,tenenbaum2020effective}. Regarding the quality of knowledge questionnaires, participants find the questions ``comprehensive and relevant'' ($M = 4.26$) and the difficulty level appropriate ($M = 3.95$). This feedback, supported by strong reliability scores, indicates that the assessments effectively aligned with the instructional content and measured participants’ comprehension.

In terms of the overall experience, the lowest-rated statement is ``The VR classroom enhanced my understanding of the material'' ($M = 3.26$), reflecting varied perceptions. This suggests that while VR tools can be effective, their success may depend on factors such as the topic, environment design, and individual learning preferences~\cite{radianti2020systematic,rojas2023systematic}. Despite this, the majority of participants expressed satisfaction with their VR classroom experience, with many agreeing to the statement, ``I would recommend VR classroom experiences to others'' (\(M = 4.16\)), indicating a generally positive perception.

Individual virtual learning environments can be designed similarly to traditional classroom settings, even in self-directed and personalized learning contexts. The integration of LLMs, which are currently highly effective and satisfactory for lecture presentations, also allows these environments to be tailored to individual needs. Incorporating peer interactions can help sustain attention and increase engagement. However, the impact of peer interactions may vary depending on the complexity of the subject matter. In more complex subjects, peer interactions are more effective in guiding attention, whereas, in less demanding subjects, their effects are less pronounced. Importantly, these interactions do not introduce excessive cognitive load and can still enhance learning outcomes.

\subsection{Limitations and Future Work}

While this study highlights the benefits of LLM-driven peer interactions, there are several limitations to consider. The sample size was relatively small, and there were gender imbalances. Expanding the range of subjects and increasing the participant pool would help strengthen the generalizability of the findings. Although avatar behaviors were animated to simulate natural interactions, the realism and quality of these animations and interactions were not formally assessed through user or expert evaluation. Future work could incorporate such assessments and further examine the quality of the AI-generated questions and answers. The study also focused on only two specific topics. Future research could explicitly investigate how topic complexity influences the suitability of virtual environments and identify which interaction settings are most effective for different content types. Additionally, examining long-term learning outcomes and the role of more active learning environments in fully LLM-driven classroom settings could offer deeper insights into how to optimize these virtual learning experiences.

\section{Conclusion}
In this study, we designed an individual learning environment with a fully LLM-driven virtual classroom, where students could interact with LLM-driven teachers and engage in a classroom setting with LLM-driven peers who also interacted with the instructor. We investigated student behavior using eye-tracking data and cognitive load assessments across two interaction conditions: one where LLM-driven peers asked questions and interacted with the teacher and another where peer interactions were not present. Our findings reveal that LLM-driven peer interactions significantly enhanced student engagement and attention, particularly in complex subjects like the Double-Slit Experiment. The interaction of LLM-driven peers increased the duration of students' fixated time on the primary instructional content, promoting sustained attention and improving learning outcomes. Although cognitive load increased in complex subjects, this was primarily attributed to the heightened attention participants directed toward the learning material. In less complex subjects, peer interactions did not increase attention or cognitive load, yet they demonstrated the potential to enhance learning outcomes without introducing excessive cognitive load.

LLM-driven peer interactions in virtual learning environments not only replicate real-world classroom dynamics but also have the potential to improve learning by keeping students engaged and focused for longer periods. This approach could be particularly valuable in higher education or specialized training, where understanding difficult concepts is crucial. Additionally, incorporating LLM-driven peers allows educators to create more personalized and interactive virtual learning environments, making advanced educational opportunities accessible to a broader range of learners. These insights may also support the development of self-directed learning environments by helping to sustain learner attention and provide more effective, personalized learning experiences. Future research should focus on expanding these insights by exploring more diverse subject areas, different types of interactions, and the long-term impact on learning retention, to fully understand the potential of LLM-driven classrooms in supporting personalized and active learning experiences.

%
\bibliographystyle{abbrv}
\bibliography{sigproc}  
%

\balancecolumns
\end{document}